# Eurasian Cooling Patterns in the CMIP5 Climate Models

Stephen Outten[1], Richard Davy[1], and Linling Chen[2]

[1] Nansen Environmental and Remote Sensing Center, Bjerknes Centre for Climate Research, Bergen, NORWAY

[2] Department of Earth Science, University of Bergen, Bergen, NORWAY

Email: stephen.outten@nersc.no

Tel. +47 4542 5074

Keywords: Eurasian cooling, CMIP5 Models, temperature trends, teleconnections, Arctic sea ice

# Abstract

The Arctic has warmed dramatically compared to the global average over the last few decades. During this same period, there have been strong cooling trends observed in the wintertime, near-surface air temperature over central Eurasia, a phenomenon known as Eurasian cooling. Many studies have suggested that the loss of sea ice, especially in the Barents and Kara Seas, is related to the cooling over Eurasia, although this connection and its possible mechanism is still a source of heated debate. Observations and reanalyses show a clear pattern of co-variability between Arctic sea ice and Eurasian wintertime temperatures. However, there is an open question regarding how robustly this teleconnection pattern is reproduced in the current generation of climate models.

This study has examined Eurasian cooling in twenty models from the Coupled Model Intercomparison Project Phase 5 (CMIP5), both in terms of temperature trends and by using singular value decomposition to identify patterns of co-variability between Arctic sea ice concentrations and near-surface temperatures over Eurasia. These are compared to trends and patterns of co-variability found in the ERA-Interim reanalysis. While most of the CMIP5 models have a robust pattern of co-variability that is similar to that found in the reanalysis, many fail to reproduce the wintertime cooling that has been observed. An examination of Arctic sea ice extent in the models suggests one possible explanation is the inability of the models to accurately simulate the wintertime sea ice changes over critical locations such as the Barents and Kara Seas.

# 1. Introduction

The Arctic has warmed more than twice as fast as the global average since the mid 20th century, a phenomenon known as Arctic amplification, AA (Serreze et al., 2009; Cohen et al., 2014). The AA signal has increased since late 1990s, showing a warming rate of more than six times the global average (Huang et al., 2017). At the same time, there have been cooling trends in near surface air temperature in the Northern Hemisphere continents, e.g. the US (Cohen et al., 2018a) and Eurasia (McCusker et al., 2016), in the last decades. This is the so-called 'Warm Arctic and Cold Eurasia' – WACE pattern (Overland et al., 2011; Cohen et al., 2014; Kug et al., 2015). In fact, both 1920-1940 and 1990-2010 periods are characterized by this temperature anomaly pattern (Chen et al., 2018; Wegmann et al., 2018), but the causes of this pattern are still under debate.

Some studies suggested a connection to the Arctic sea-ice decline (Honda et al., 2009; Petoukhov and Semenov, 2010; Liu et al., 2012; Outten and Esau, 2012; Mori et al., 2014; Kug et al., 2015; Semenov, 2016; Cohen et al., 2018; Wegmann et al., 2018). Regionally, the increased heat and moisture are transported over adjacent continents, increasing the air temperature and precipitation in these regions (Screen et al. 2015; Deser et al. 2010). The increased Arctic temperature also reduces the meridional temperature gradient, which weakens the tropospheric westerly winds of the jet stream (Deser et al. 2010; Sun et al. 2015). Furthermore, reduction in the Arctic summer-to-fall sea ice extent and increased winter snow cover over Eurasian continent during recent decades has led to an intensified Siberian High (Honda et al., 2009; Mori et al., 2014) and a negative phase shift of the Arctic Oscillation and North Arctic Oscillation (Liu et al., 2012; Nakamura et al., 2015). This may in turn have resulted in greater meandering of NH jets (Francis and Vavrus 2015; Cattiaux et al., 2016; Di Capua and Coumou, 2016; Vavrus et al., 2017), which often brings anomalously cold weather to East Asia.

AA has also enhanced stratosphere-troposphere coupling through the vertical propagation of planetary-scale waves in late fall and early winter, and often leads to a weakened or geographically shifted stratospheric polar vortex (Kim et al., 2014; Zhang et al., 2016; Nakamura et al., 2016a; Kretschmer et al., 2017), which can affect surface weather via downward wave propagation. However, other studies find no significant contributions from sea ice (Woollings et al., 2014; Li et al., 2015; McCusker et al., 2016; Sun et al., 2016). It is a major challenge to quantify or distinguish the impact of Arctic forcing amid the substantial natural variability and/or signal from the tropics, combined within the short time span of clear Arctic warming signal (Cohen et al., 2014; Overland et al., 2016; Francis et al., 2017).

While the debate over the mechanism goes on, there is an open question about the capabilities of the climate models to reproduce this WACE teleconnection pattern (Mori et al. 2014; Outten et al. 2013). This paper investigates the representation of Eurasian wintertime cooling in 20 CMIP5 models compared to reanalysis and examines the patterns of co-variability between surface air temperature and sea-ice concentration. Section 2 outlines the data sources and methods used in this study, while Section 3 presents the surface temperature trends and patterns of co-variability as they appear in the ERA-Interim reanalysis. Section 4 investigates the temperatures in the CMIP5 models including the wintertime temperature trends, while Section 5 explores the patterns of co-variability in the CMIP5 models. Discussion and conclusions are given in Section 6.

## 2. Data Sources and Analysis

The European Centre for Medium Range Forecasting's reanalysis, ERA-Interim, was used in this work on a grid of 0.5° horizontal resolution with daily temporal resolution (Dee et al. 2011). The meteorological parameters examined are the 2-meter surface air temperature and the sea ice concentration. ERA-Interim was selected due to its high spatial resolution and because it covers the period of interest, from 1989-2010. This period was chosen to maintain inter-comparability with previous studies which examined Eurasian cooling in five historical CMIP5 simulations (Outten and Esau, 2012; Outten et al. 2013). This period also has the advantage of not including the 1970's, which often introduces problems for calculating long-term trends due to the major changes to the global observing system made during those years (Screen and Simmonds, 2011).

The model data used in this work comes from twenty atmosphere-ocean global climate models (AOGCMs) of Phase Five of the Coupled Model Inter-comparison Project (CMIP5). The Historical experiment of CMIP5 included historical estimates and observations of natural and anthropogenic forcings and covered the period of 1850 to 2005, while the future experiment used forcings from the Representative Concentration Pathway (RCP) 4.5 and covered the period of 2006 until 2100. Data from these two experiments was combined to create continuous time series for the period of interest. Model data was also used from the equivalent Atmospheric Model Intercomparison Project (AMIP) experiment of CMIP5 where available. This was the atmosphere-only component of the coupled models constrained with realistic sea ice concentration and sea surface temperature, following the Atmospheric Model Inter-comparison Project (AMIP) protocol (Gates, 1992). Only seventeen of the twenty models had AMIP experiments available. The three models without compatible AMIP runs were the FIO-ESM, HadGEM2-AO, and CESM1-CAM5 models. The design of these experiments is covered in more detail in Taylor et al. (2012). The names of the twenty models selected along with the modeling groups who used them are given in Table 1. To allow for inter-comparison between the models and between the atmospheric and surface fields, the data from all models was interpolated onto a horizontal, rectilinear grid of 1° x 1° resolution using 2D linear interpolation. Since the focus is Eurasia and the Arctic, the analysis was limited to the extratropical Northern Hemisphere ($\varphi \geq 30N$).

Trends of surface air temperature were calculated from the monthly anomalies created by the subtraction of the climatological mean for each month for the period under analysis. The monthly anomalies from December, January and February were used to calculate a wintertime temperature trend by applying a linear fit to these anomalies. These trends were filtered for statistical significance at the 95% level against the null hypothesis of the hemispheric average annual trend.

Singular Value Decomposition (SVD) is used to assess the co-variability of the surface temperature with the sea ice concentration over the Northern Hemisphere. By correlating the primary mode of co-variability against the temperature and sea ice concentration, homogeneous correlation maps were created to elucidate the pattern of co-variability associated with the primary mode. This was done for both reanalysis and the 20 CMIP5 models. SVD is a statistical method for identifying co-variability and does not imply causality.

## 3. Eurasian Cooling Patterns in Reanalysis

The wintertime surface air temperature trends in ERA-Interim show regions of strong warming over Western Greenland and the northern Barents Sea, between Svalbard and Northern Russia (Figure 1). The strongest warming is found over Southern Greenland, where it peaks at 6.2 Kelvin per decade (K dec$^{-1}$). In the region of warming north of Eurasia however, the warming reaches a maximum trend of 5.0 K dec$^{-1}$. While these trends are large, they are highly localized and are comparable to trends found in previous studies (Mori et al. 2014; Outten et al. 2013) and compared to observations from some of the few stations in the region (Outten and Esau, 2012, supplemental). Figure 1 also shows a region of strong cooling over central Eurasia, and a second region of weaker cooling centered over Scandinavia, a pattern consistent with our previous works (Outten and Esau, 2012; Outten et al. 2013). The Eurasian cooling in ERA-Interim for this period peaks at -4.9 K dec$^{-1}$, at approximately the same location as negative temperature anomalies observed in the NCEP/NCAR reanalyses (Overland et al. 2008; Semenov and Latif, 2015), minimum in temperature trends from satellite measurements (Comiso et al. 2003), and similar negative temperature trends found in observations from the HadCRUT4 dataset (Kug et al. 2015). This pattern of wintertime warming in the Arctic with associated cooling over Eurasia is now well established in the literature and the results in Figure 1 are consistent with those studies.

Although the mechanistic links between the Arctic warming and Eurasian cooling are still highly debated (Honda et al., 2009; Petoukhov and Semenov, 2010; Mori et al. 2014; Francis and Vavrus 2015; McCuscker et al. 2016; Blackport et al. 2019; Mori et al. 2019), many studies have suggested links to the loss of Arctic sea ice. Using Singular Value Decomposition analysis, we are able to investigate the co-variability between surface air temperatures and sea ice concentrations in ERA-Interim (Figure 2). The primary mode of co-variability shows that sea ice reduction, especially over the northern Barents Sea/Kara Sea region, is accompanied by increased air temperatures over this region of the Arctic and by reduced air temperatures over Central Eurasia. As this is a pattern of variability, the reverse is true when the sea ice increases. These variations in air temperature are significant at the 95% level and this WACE pattern is well established in the literature (Cohen et al., 2014; Chen et al., 2018; Mori et al. 2014; Kug et al. 2015; Outten and Easu, 2012; Wegmann et al., 2018). The pattern is also found in other reanalyses, e.g. NCEP/NCAR, ERA20C, CERA20C and 20CRv2C (Outten et al. 2013; Chen et al., 2018). It should be noted that SVD is a statistical methodology and does not in any way demonstrate a physical connection between the variations. In this case, it shows only that the variations in sea ice are accompanied by variations in temperature over both the Arctic and over Central Eurasia.

The region of positive temperature variation over the northern Barents Sea is collocated with the warming in this region, as shown in Figure 1. Similarly, the region of negative temperature variation over Central Eurasia is collocated with the major region of cooling shown in the Figure 1. Perhaps more interesting though are the locations where Figures 1 and 2a do not agree. Despite there being a region of strong cooling over Scandinavia, there is no associated negative temperature variability shown in Figure 2a. This suggests that while the Eurasian cooling may be associated with sea ice loss in the Barents/Kara Sea region, the cooling over Scandinavia is not. Studies have suggested that the loss of sea ice and the associated local warming may alter the atmospheric flow extending over Eurasia, perhaps by triggering a stationary Rossby wave train or perhaps by increased frequency of Eurasian blocking events (Honda et al. 2009; Semenov and Latif, 2015; Mori et al, 2015; Zappa et al. 2018). Such a mechanism would account for the changes in sea ice concentration in the northern Barents/Kara Sea region having little to no impact on the temperatures over Scandinavia, and hence the lack of associated temperature variability found in Figure 2a. Furthermore, while some positive temperature variability is found over Southern Greenland, it would not account for the strong warming seen in this region in Figure 1. Again, since this location is upstream of the Barents/Kara Sea where the sea ice variability is largest in the primary mode, it is not expected that such sea ice loss would be associated with temperatures over Southern Greenland. It has also been suggested that rapid warming in Western Greenland and the Canadian Arctic Archipelago, similar to that seen in ERA-Interim, is driven by unforced natural variability relate to the tropics (Ding et al. 2014).

## 4. Temperature trends in CMIP5 models

While the previous section has shown the trends in surface temperatures and the pattern of covariability associated with Eurasian cooling in the ERA-Interim reanalysis, the question is how well those trends and patterns are reproduced by our current generation of climate models. Before addressing that question though, some initial investigation of the models' capabilities is required.

As part of the initial exploratory data analysis, the spread in temperature between the models was examined in comparison to ERA-Interim using Taylor diagrams (Figure 3). The coupled models are all broadly clustered around an RMS of 0.5 K compared to ERA-Interim, with the CSIRO-Mk3.6.0 model showing the best agreement to the reanalysis. The similarity in RMS between the models is likely related to the fact that all models are tuned to minimize RMS error in several key parameters including surface air temperature (Schmidt et al. 2017). However, the models show a wide-spread in the standard deviation compared to ERA-Interim's 0.49 K, and a wide range of pattern correlations to ERA-Interim. Eleven of the models have a lower standard deviation, indicating a smaller amplitude in the temperature variations in the extratropical northern hemisphere. The INM-CM4 model shows the narrowest spread in temperatures with a standard deviation of only 0.33 K. This model also has the lowest pattern correlation with R=0.13. The other nine models overestimate the spread in temperatures, with the greatest standard deviation of 0.61K found in the HadGEM2-AO model, though this also has the highest pattern correlation at R=0.65. The high spread in temperatures appears to be a property of the HadGEM2 model since HadGEM2-ES, which is an alternate version of the same model, has a comparable standard deviation of 0.60 K. This also has a high pattern correlation at R=0.62. The best agreements with the reanalysis in terms of spread of temperatures are found in the GISS-E2-R and BCC-CSM1.1, having standard deviations of 0.49 and 0.50 K respectively.

This large spread in standard deviations of temperatures across the models raises concerns about their ability to accurately reproduce the Eurasian cooling pattern. The underlying hypothesis from literature proposes that during the wintertime, changes in sea ice concentration, especially in the Barents/Kara Sea regions, are related to changes in the near surface air temperatures of central Eurasia. This means that the cooling is either a response to the change in sea ice concentration, or that it is a response to some unknown factor to which the sea ice concentration is also reacting. Therefore, in order for the models to reproduce the Eurasian cooling pattern, they would need to have some consistency in their reaction to changes found in the sea ice, which does not appear to be the case. However, the fully coupled models all have very different sea ice concentrations, so in order to assess how consistently the models respond to changes in sea ice, we examined the temperature in the AMIP experiments for seventeen of the same twenty models that were available (Figure 3). Since the AMIP experiments use realistic sea ice concentration and sea surface temperatures, the atmospheric components of the fully coupled models are all constrained with the same ocean forcings. The models are now tightly grouped in the Taylor diagram around the standard deviation of 0.49 K found in ERA-Interim, with all values lying between 0.4 and 0.5 K. This demonstrates that the atmospheric components of the fully coupled climate models are consistently responding to the changes in the ocean surface, and therefore they should be capable of reproducing the pattern of variability associated with Eurasian cooling.

Before examining the pattern of co-variability in the models, we first examine the trends in surface air temperature to see which of the twenty models show some Eurasian cooling (Figure 4). The trends in the twenty models are shown on the same scale as used in Figure 1 for ease of comparison with the trends in ERA-Interim. Many of the models do not show any significant Eurasian Cooling. Two models show strong cooling over Central Eurasia (CanEMS2, NorESM1-M), while three more show some weak cooling (CNRM-CM5, MPI-ESM-MR, MRI-CGCM3), and four models show cooling over Europe (ACCESS1.3, EC-Earth, FGOALS-g2, CESM1-CAM5). No model shows cooling over both Central Eurasia and Europe, as seen in the reanalysis. Both of the models that have strong Eurasian cooling, also have strong warming over the Barents Sea, indicative of sea ice reduction. Of all the models, the temperature trends in CanESM2 most closely resemble those found in ERA-Interim, while the most different trends are found in either FIO-ESM or MIROC-ESM. Both of these models show strong wintertime warming over Central Eurasia along with strong cooling over Greenland and some cooling over the Barents Sea, which are all opposite of the trends found in ERA-Interim. Based on Figure 4 it is clear that in general the current generation of climate models do not reproduce the Eurasian wintertime cooling as found in reanalyses and observations. However, if there were some natural variability within Earth's climate system that is not related to forcings provided to the model, we would not expect the models to reproduce the Eurasian wintertime cooling over the same period as it is observed in reanalyses and observations.

## 5. Patterns of co-variability in CMIP5 models

We now investigate the robustness of teleconnection patterns associated with Eurasian cooling in the twenty climate models by using SVD analysis to examine the co-variability between sea ice concentration and near surface air temperature (Figure 5). The scale is consistent with Figure 2 which showed the co-variability derived for the ERA-Interim reanalysis, allowing for direct comparison. However, SVD analysis, like Empirical Orthogonal Function (EOF) analysis, does not always produce a clear separation between the different modes of variability. If the separation of eigenvalues for the modes is small, then the modes are unlikely to represent significant and unique spatio-temporal patterns, and therefore they must be treated as a coupled pair, rather than individual modes. Table 2 shows the co-variance for the first and second mode of co-variability in the twenty CMIP5 models, and reveals that 4 models have small separations in their eigenvalues. These are BCC-CSM1.1, ACCESS1.3, CSIRO-Mk3.6.0 and the FIO-ESM models. In all four of these models, the second mode of co-variability was found to more closely resemble the Eurasian cooling pattern and hence are shown in Figure 5 in place of the primary mode. However, the first and second mode of co-variability for all 20 models are shown in Supplemental Figures S1 and S2 respectively.

Thirteen of the models show a pattern of co-variability similar to that found in the reanalysis, with the critical features of reducing sea ice concentrations mainly focussed in the Barents Sea, being associated with warming over the Barents Sea and near Arctic Ocean, and with cooling over Central Eurasia. As this is a pattern of variability, the reverse is also true, i.e. increasing sea ice in the Barents Sea is associated with cooling locally and warming over Central Eurasia in these same thirteen models. Four more models also show patterns that are partially consistent with that found in ERA-Interim.

The HadGEM2-ES model has sea ice reduction focussed over the eastern Barents Sea which is associated with warming limited to the local region and not extending up to the Arctic Ocean, and cooling which is displaced southwest towards the Middle East and Mediterranean. This differs from the other HadGEM2 model, HadGEM2-AO, which shows sea ice reduction focussed north of the Barents Sea being associated with co-located warming and with very weak cooling over Central and South Eastern Eurasia. The location of the cooling and the intense warming over the high Arctic Ocean are broadly similar to what is found in the GISS-E2-R model, which shows weaker sea ice reduction that is not focussed in any specific location, associated with broad warming over the Arctic Ocean and weak cooling displaced southeast towards Japan. Finally, the CESM1-CAM5 model shows sea ice reduction focussed on the Barents Sea associated with very little local warming. Instead, the warming is located over the Arctic Ocean north of the Laptev and Eastern Siberian Seas, while the cooling extends from Central Eurasia towards Eastern Europe. While all four of these models show differences in their pattern of co-variability from what is seen in the reanalysis, they all have a primary mode showing a reduction of sea ice north of Eurasia, mostly over the Barents Sea, associated with warming over the Arctic, and weak but significant cooling somewhere over the Eurasian continent.

Only three of the twenty models showed a pattern of covariance that was very different from that found in the reanalysis. These were the ACCESS1.3, CSIRO-Mk3.6.0, and the FGOALS-g2 models. ACESS1.3 and CSIRO-Mk3.6 are variants of the same model and in both, the SVD was unable to separate the primary and secondary modes of co-variability. The pattern of increased sea ice in the central/western Barents Sea and Kara Sea being associated with warming over much of Greenland and cooling over Europe and Central Eurasia is broadly consistent between these two models. While the cooling extends from Central Eurasia to Europe in the ACCESS1.3 model, this is also true for the CSIROmk3.6.0. model, it is just not significant and hence not shown. The main difference occurs in the Greenland Sea where ACCESS1.3 shows increased sea ice and CSIRO-Mk3.6.0 shows decreased sea ice. Given that these two models have the weakest separation of eigenvalues (Table 2), it is questionable as to whether or not these patterns can be considered reliable.

The third model which shows a distinctly different pattern of co-variability from the reanalysis is the FGOALS-g2 model. This shows an extremely weak variation in sea ice concentrations with no single obvious centre of variation. The sea ice changes are apparently associated with cooling over Europe, extending out over the Nordic Sea and east to Central Eurasia. Investigation of this models revealed that FGOALS-g2 has the lowest sea ice extent and by far the smallest variability in sea ice concentration of all twenty models, both in terms of long-term change over the whole of the 20$^{th}$ century and in terms of interannual variability (not shown). Given the lack of realistic variations in the sea ice concentrations in this model, it is unsurprising that it cannot reproduce the pattern of co-variability associated with Eurasian cooling.

In summary, most of the CMIP5 models studied here appear to robustly reproduce the pattern of co-variability between near surface air temperatures and sea ice concentrations as found in reanalysis. All but three of the models show a decrease in sea ice concentration, usually greatest in the Barents Sea, associated with warming over the Barents Sea and Arctic Ocean and with cooling over Central Eurasia. The three models which failed to reproduce this pattern either could not have their primary and secondary modes of co-variability separated by the SVD analysis or had unrealistically small variabilities in their sea ice concentrations. This suggests that the current generations of climate models should be capable of reproducing the Eurasia cooling as it relates to changes in sea ice concentration and gives some guide as to which models are most successful at reproducing this teleconnection pattern.

## 6. Discussion and Conclusions

The work presented in the previous section shows that in general, the current generation of climate models robustly reproduce the teleconnection pattern associated with Eurasian cooling, that is they show variations in the sea ice concentrations, especially over the Barents/Kara Sea region, are accompanied by changes in the near surface air temperature over Central Eurasia. However, the trend analysis shows that the models are not reproducing Eurasia cooling, that is they do not show negative temperature trends in Central Eurasia. This suggests two possibilities, either the temperatures over Central Eurasia are being dominated by other factors which are overriding this pattern of co-variability, or the models are poorly reproducing the changes in sea ice concentration, especially over the Barents/Kara Sea region.

Regarding the first possibility, it should be noted that in the models which show this pattern of co-variability, it is always the primary mode, or part of a coupled pair including the primary mode (which occurs where the separation of eigenvalues is insufficient for the modes to be distinct, as in the case of the BCC_CSM1.1 and FIO-ESM models). Table 2 indicates that the percentage of covariance explained by this primary mode is between 30% and 50% in most models, with 6 models showing higher covariance. Only two models show covariance below 30% and these are ACCESS1.3 and CSIRO-Mk3.6.0 models; both variants of the same model and neither showing the Eurasian cooling teleconnection pattern (Figure 5). Beyond this, Table 2 also indicates that the strength of the coupling between sea ice concentration and surface air temperatures is strong and highly significant in all models for the primary mode. This is given the by the high correlation coefficients, which range from R=0.62 to R=0.82. These values are comparable to those found in the SVD analysis for ERA-Interim, which has a primary mode co-variability of 50.7% and correlation of R=0.79. The strength and significance of the co-variations in the models suggest that there are clear variations in the sea ice concentration which are well correlated by variations in the temperature over Central Eurasia. This does not eliminate the possibility that the temperatures in Central Eurasia are dominated by other factors, and indeed this is very possible, however given the strength of these co-variances it is perhaps unexpected that so few of the models show any significant cooling over Central Eurasia.

We now consider the second possibility that the models are struggling to reproduce the changes in the wintertime sea ice concentrations. Previous studies have linked the Eurasian cooling to sea ice removal over the Barents and Kara Seas (Honda et al., 2009; Liu et al., 2012; Outten and Esau, 2012; Mori et al., 2014; Kug et al., 2015; Semenov, 2016; Cohen et al., 2018; Wegmann et al., 2018). This suggests that it is not only the quantity of Arctic sea ice that influences Eurasian cooling, but also the location in which the sea ice reduction occurs that is important. A model could, for example, accurately reproduce the total wintertime Arctic sea ice extent, yet still fail to reproduce the regional changes over the critical locations necessary to contribute to Eurasia cooling. The Barents and Kara Sea are regions where the models struggle to accurately reproduce the sea ice extent when compared to reanalysis (Figure 6), with some models having the Barents Sea ice free in the winter, while others have sea ice extending across the entire Barents Sea. Thus, while variations in sea ice concentration in this region may be well correlated with variations in temperature over Central Eurasia, these variations may be occurring against a background of almost complete ice cover or almost complete open water, with little to no longer term trend. Figure 6 shows only the mean DJF sea ice extent in the models compared to the reanalysis and further work would be required to assess how these sea ice extents vary from year to year, however, this does demonstrate that the climate models are struggling to reproduce the regional changes in sea ice concentrations in this critical region.

It is important to reiterate that the SVD analysis presented here does not demonstrate that changes in sea ice concentration result in changes in temperatures over Eurasia. The SVD analysis only shows that changes in sea ice are co-occurring with such temperature changes. It may be that sea ice concentration and the temperatures over Eurasia are independently affected by a third property, such as large-scale atmospheric flow, or it may be that changes in sea ice concentration reinforce and contribute to atmospheric circulation anomalies which influence the temperatures over Central Eurasia, as suggested by Semenov and Latif (2015). In either case, the SVD analysis would only show that the sea ice concentrations and the temperatures over Eurasia are co-varying.

This work has examined the representation of Eurasian cooling in the current generation of climate models. The pattern of co-variability found in the ERA-Interim reanalysis, and in other reanalyses (Outten et al. 2013), associated with Eurasian cooling is well reproduced in many of the CMIP5 climate models. Despite this robust pattern of co-variation between the sea ice concentration and temperatures over Central Eurasia, most of the models fail to reproduce the significant wintertime cooling that has been observed. One possible explanation for this may lie in the models' capabilities to accurately simulate the wintertime sea ice changes over critical locations such as the Barents and Kara Seas. Further work is needed to demonstrate the impact of the accurate location of sea ice change on the reproduction of Eurasian cooling. It is hoped that this work may provide some basis for model selection for future studies of Eurasian cooling. Furthermore, given that the major dynamical features for changing the NH mid-latitude weather, such as storm tracks, the jet stream and atmospheric blocking are still poorly simulated by many models due to biases in sea surface temperature, low horizontal resolution and/or missing orographic drag (Anstey et al., 2013; Masato et al., 2013; Zappa et al., 2013; Shepherd, 2014; Wang et al., 2014; Woollings et al., 2014; Zappa et al., 2014; Yim et al., 2015; Pithan et al., 2016; Davini et al., 2017; Mitchell et al., 2017), it is possible that sea-ice influence on the mid-latitude weather will become more apparent as climate models improve.

# Acknowledgement

*This work was funded by the EuRCool project funded by the Bjerknes Centre for Climate Research (BCCR), by the KEYCLIM project funded by grant number 295046 and EuropeWeather/ (231322/F20) project funded by the Norwegian Research Council. We acknowledge the World Climate Research Programme's Working Group on Coupled Modelling, which is responsible for CMIP, and we thank the climate modeling groups (listed in Table 1 of this paper) for producing and making available their model output. For CMIP the U.S. Department of Energy's Program for Climate Model Diagnosis and Intercomparison provides coordinating support and led development of software infrastructure in partnership with the Global Organization for Earth System Science Portals.*

Tables

| ID | Modelling Centre | Institute ID | Model Name |
|---|---|---|---|
| A | Beijing Climate Center, China Meteorological Administration | BCC | BCC-CSM1.1 |
| B | Canadian Centre for Climate Modelling and Analysis | CCCMA | CanESM2 |
| C | Centre National de Recherches Météorologiques / Centre Européen de Recherche et Formation Avancée en Calcul Scientifique | CNRM-CERFACS | CNRM-CM5 |
| D | Commonwealth Scientific and Industrial Research Organization (CSIRO) and Bureau of Meteorology (BOM), Australia | CSIRO-BOM | ACCESS1.3 |
| E | Commonwealth Scientific and Industrial Research Organization in collaboration with Queensland Climate Change Centre of Excellence | CSIRO-QCCCE | CSIRO-Mk3.6.0 |
| F | The First Institute of Oceanography, SOA, China | FIO | FIO-ESM |
| G | EC-EARTH consortium | EC-EARTH | EC-EARTH |
| H | Institute for Numerical Mathematics, Russia | INM | INM-CM4 |
| I | Institut Pierre-Simon Laplace | IPSL | ISPL-CM5A-MR |
| J | LASG, Institute of Atmospheric Physics, Chinese Academy of Sciences | LASG-CESS | FGOALS-g2 |
| K | Japan Agency for Marine-Earth Science and Technology, Atmosphere and Ocean Research Institute (The University of Tokyo), and National Institute for Environmental Studies | MIROC | MIROC-ESM |
| L | Met Office Hadley Centre | MOHC | HadGEM2-ES |
| M | Max-Planck-Institut für Meteorologie (Max Planck Institute for Meteorology) | MPI-M | MPI-ESM-MR |
| N | Meteorological Research Institute | MRI | MRI-CGCM3 |
| O | NASA Goddard Institute for Space Studies | NASA GISS | GISS-E2-R |
| P | National Center for Atmospheric Research | NCAR | CCSM4 |
| Q | Norwegian Climate Centre | NCC | NorESM1-M |
| R | National Institute of Meteorological Research/Korea Meteorological Administration | NIMR/KMA | HadGEM2-AO |
| S | NOAA Geophysical Fluid Dynamics Laboratory | NOAA GFDL | GFDL-ESM2M |
| T | Community Earth System Model Contributors | NSF-DOE-NCAR | CESM1-CAM5 |

Table 1. List of CMIP5 models used in this study, along with the respective modelling centre, group, or consortium that ran the models and prepared the output for the CMIP5 archive.

| ID | Model Name | 1st Mode Covariance | 2nd Mode Covariance | 1st Mode R (p≤0.05) |
|---|---|---|---|---|
| A | BCC-CSM1.1 | 31.0% | 24.6 % | 0.76 |
| B | CanESM2 | 47.9% | 16.6 % | 0.73 |
| C | CNRM-CM5 | 72.5% | 7.7 % | 0.64 |
| D | ACCESS1.3 | 27.5% | 22.4 % | 0.80 |
| E | CSIRO-Mk3.6.0 | 25.5% | 24.3 % | 0.62 |
| F | FIO-ESM | 31.9% | 25.5 % | 0.68 |
| G | EC-EARTH | 40.3% | 22.4 % | 0.73 |
| H | INM-CM4 | 33.9% | 13.9 % | 0.78 |
| I | ISPL-CM5A-MR | 46.5% | 18.1 % | 0.82 |
| J | FGOALS-g2 | 33.5% | 18.9 % | 0.62 |
| K | MIROC-ESM | 39.4% | 15.3 % | 0.75 |
| L | HadGEM2-ES | 30.2% | 15.6 % | 0.82 |
| M | MPI-ESM-MR | 60.5% | 13.4 % | 0.74 |
| N | MRI-CGCM3 | 59.0% | 12.4 % | 0.72 |
| O | GISS-E2-R | 54.1% | 16.3 % | 0.68 |
| P | CCSM4 | 47.9% | 18.4 % | 0.77 |
| Q | NorESM1-M | 60.8% | 12.5 % | 0.76 |
| R | HadGEM2-AO | 51.9% | 15.0 % | 0.80 |
| S | GFDL-ESM2M | 48.0% | 20.8 % | 0.78 |
| T | CESM1-CAM5 | 46.6% | 19.2 % | 0.78 |

Table 2. List of CMIP5 models with the overall co-variance between sea-ice and air temperature explained by the first and second modes, and the correlation coefficients for the first mode.

Figure Captions

Figure 1. Surface air temperature trends [K/decade] in ERA-Interim reanalysis over the period of 1989 to 2010. The plotted regions show the locations which are significant at the 95% level against the null hypothesis of the hemispheric average annual trend. Contour interval is 0.5 K.

Figure 2. Spatial distribution of the first SVD mode mapped as homogenous correlation with surface air temperature (left) and sea-ice concentration (right) from ERA-Interim. Contour interval is 0.1.

Figure 3. Taylor diagrams for northern hemisphere surface air temperature in the 20 models as compared to ERA-Interim for 1989-2010. (Left) Coupled experiments, (Right) Atmosphere-only experiments. Only 17 of the 20 models had available AMIP simulations. Contours of RMS (Green dashed) are every 0.25 K, as per standard deviation.

Figure 4. Surface air temperature trends [K/decade] in the 20 CMIP5 models over the period of 1989 to 2010. The plotted regions show the locations which are significant at the 95% level against the null hypothesis of the hemispheric average annual trend. Contour interval is 0.5 K.

Figure 5. Maps of the homogenous correlation with surface air temperature (left insert) and sea-ice concentration (right insert) for the 20 CMIP5 models of the SVD mode numbered in the bottom right corner of each plot. This is the $1^{st}$ mode of the SVD, or the $2^{nd}$ mode if there was a separation issue between the $1^{st}$ and $2^{nd}$ modes. Contour interval is 0.1.

Figure 6. Count of sea ice extent where mean DJF sea ice concentration is at least 15% in the 20 CMIP5 models. Mean DJF sea ice extent in HadISST for 1989-2010 (bold line). Contour interval is 1 model.

Figures

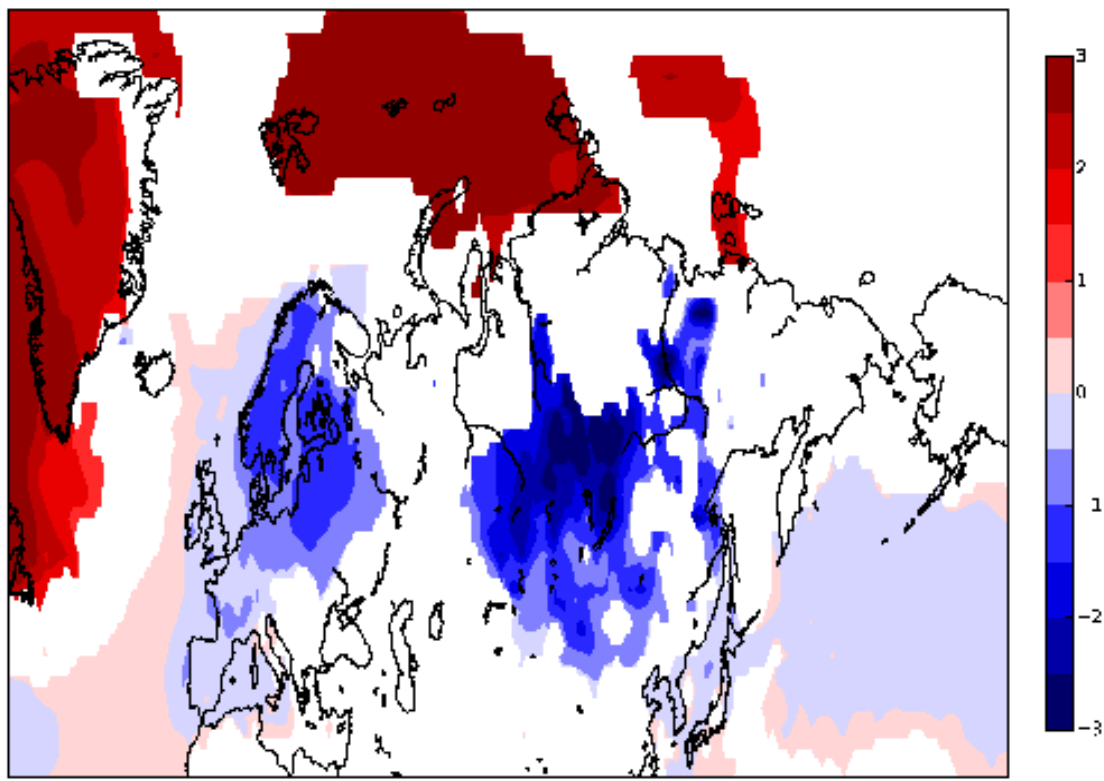

Figure 1. Surface air temperature trends [K/decade] in ERA-Interim reanalysis over the period of 1989 to 2010. The plotted regions show the locations which are significant at the 95% level against the null hypothesis of the hemispheric average annual trend. Contour interval is 0.5 K.

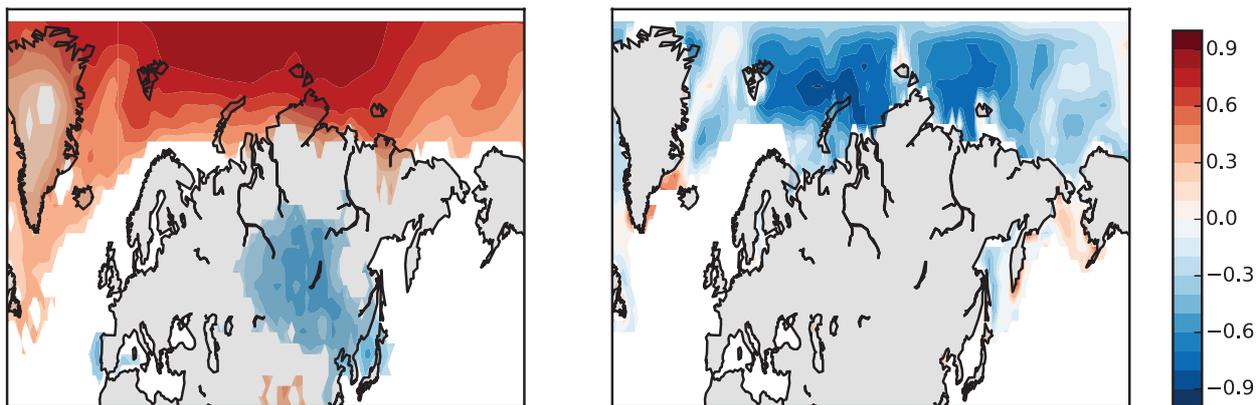

Figure 2. Spatial distribution of the first SVD mode mapped as homogenous correlation with surface air temperature (left) and sea-ice concentration (right) from ERA-Interim. Contour interval is 0.1.

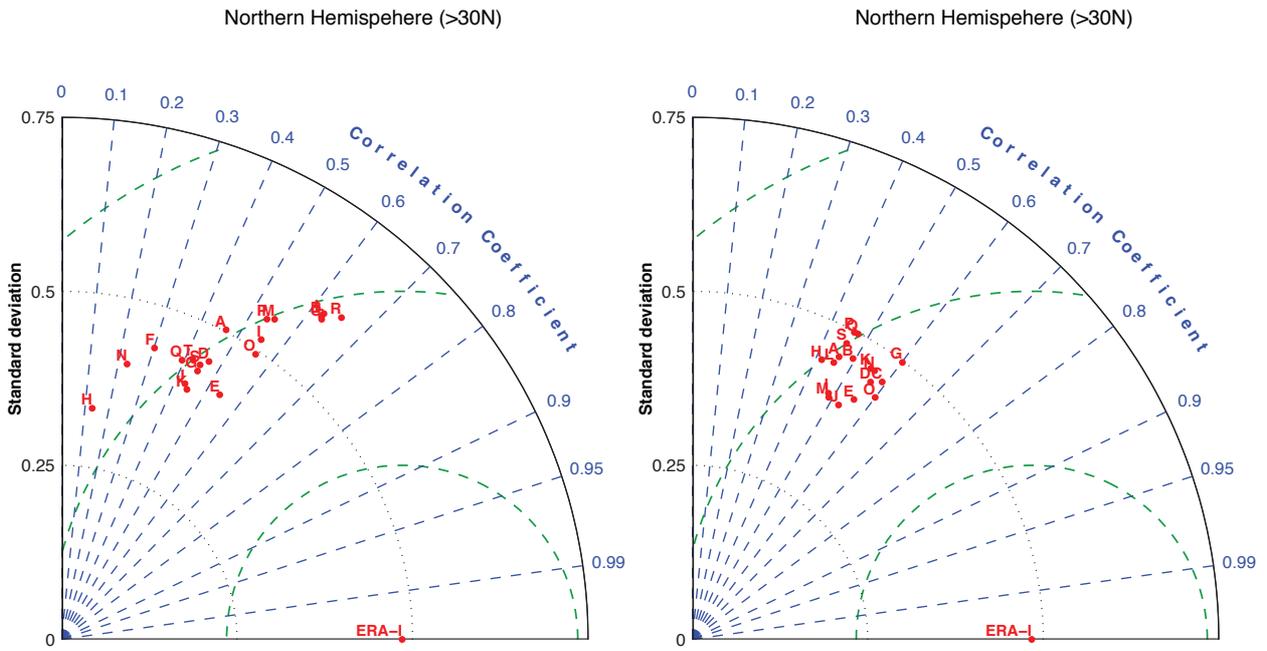

Figure 3. Taylor diagrams for northern hemisphere surface air temperature in the 20 models as compared to ERA-Interim for 1989-2010. (Left) Coupled experiments, (Right) Atmosphere-only experiments. Only 17 of the 20 models had available AMIP simulations. Contours of RMS (Green dashed) are every 0.25 K, as per standard deviation.

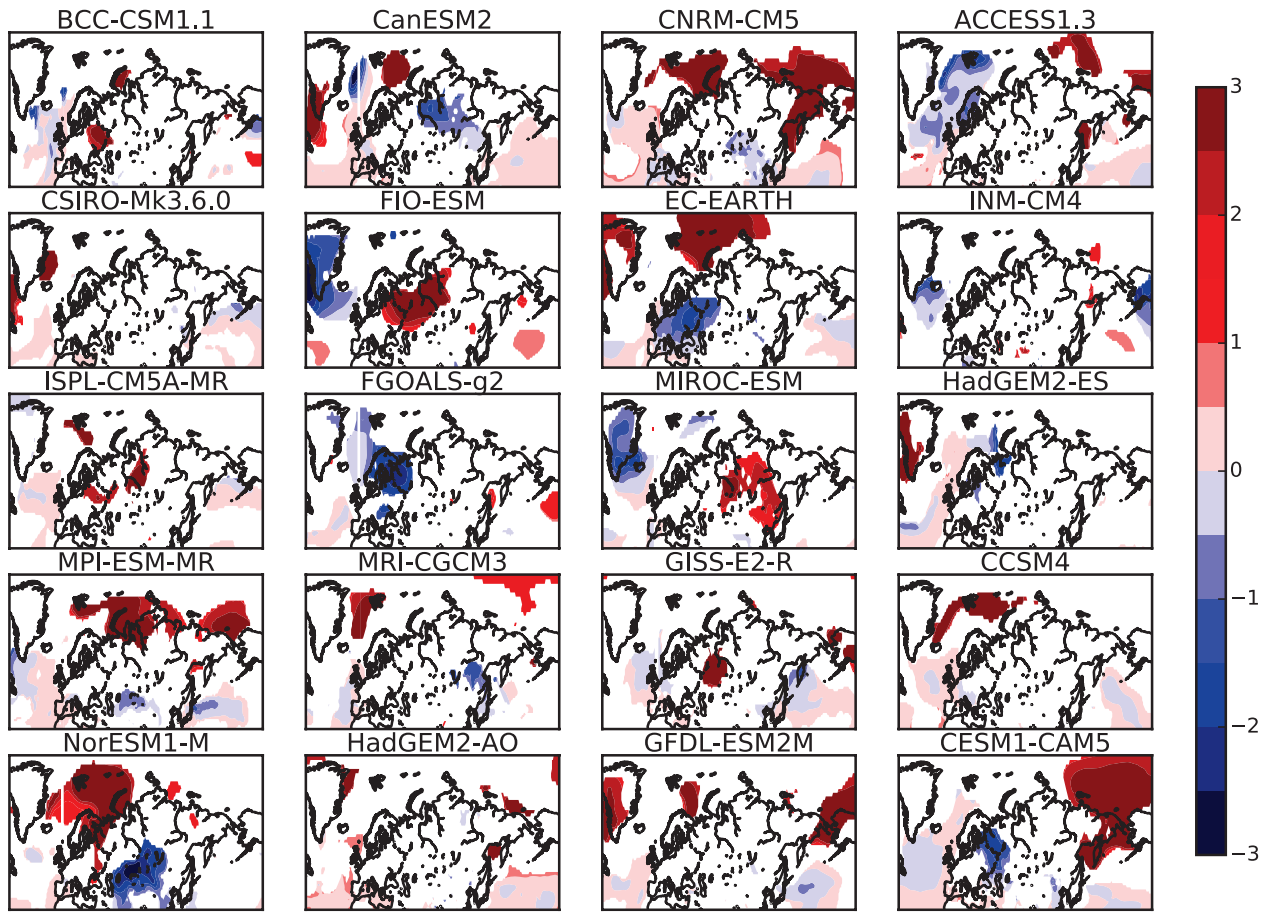

Figure 4. Surface air temperature trends [K/decade] in the 20 CMIP5 models over the period of 1989 to 2010. The plotted regions show the locations which are significant at the 95% level against the null hypothesis of the hemispheric average annual trend. Contour interval is 0.5 K.

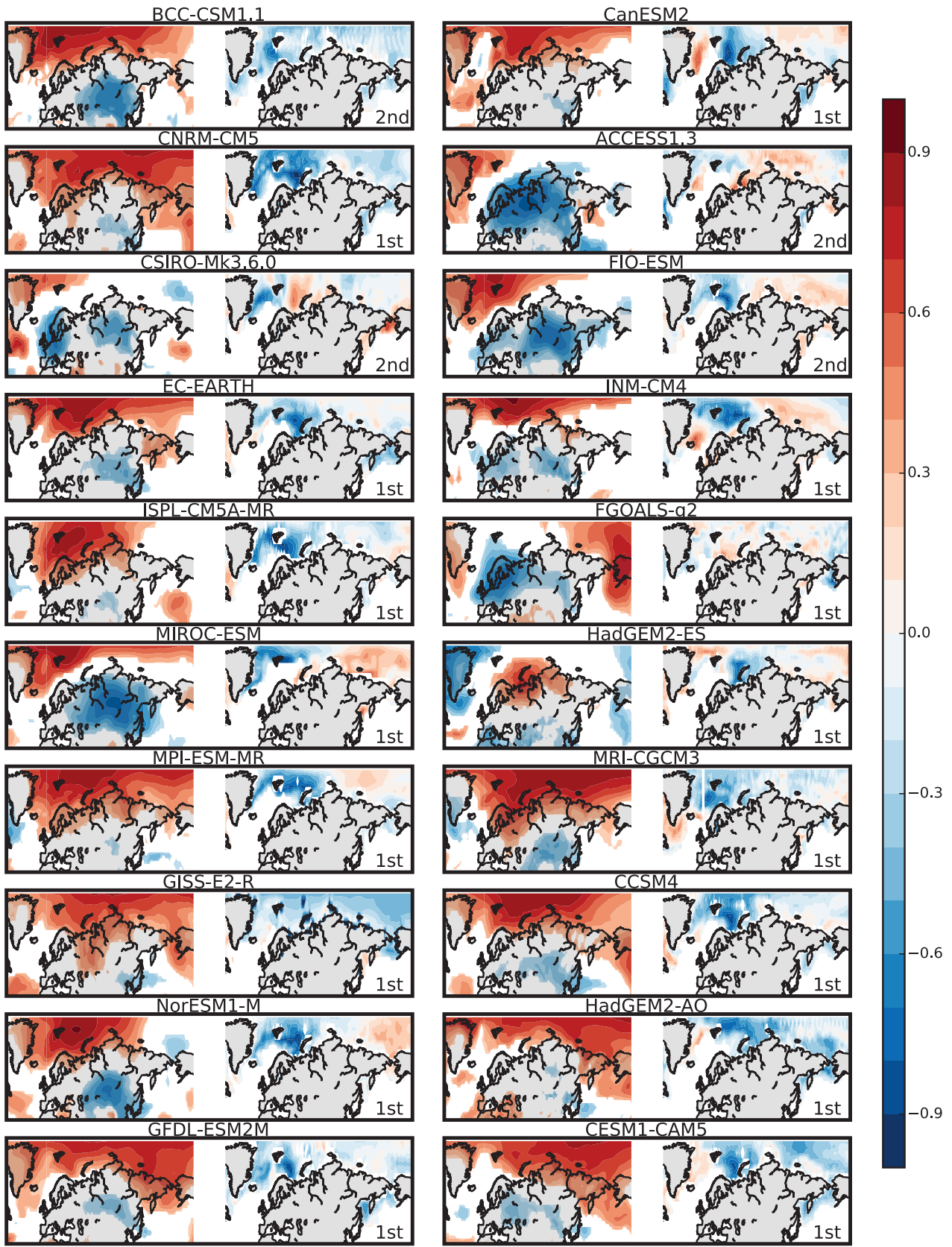

Figure 5. Maps of the homogenous correlation with surface air temperature (left insert) and sea-ice concentration (right insert) for the 20 CMIP5 models of the SVD mode numbered in the bottom right corner of each plot. This is the 1st mode of the SVD, or the 2nd mode if there was a separation issue between the 1st and 2nd modes. Contour interval is 0.1.

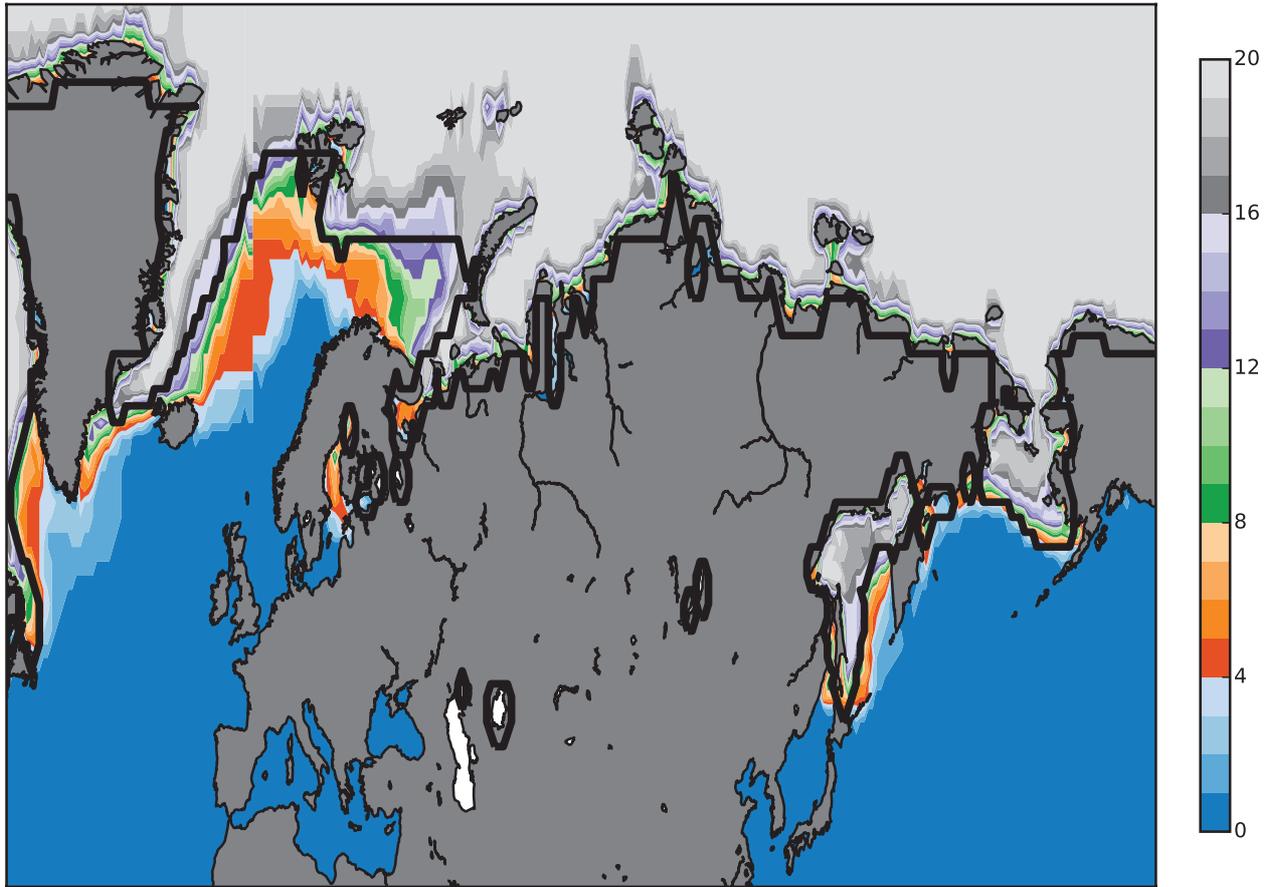

Figure 6. Count of sea ice extent where mean DJF sea ice concentration is at least 15% in the 20 CMIP5 models. Mean DJF sea ice extent in HadISST for 1989-2010 (bold line). Contour interval is 1 model.